# Suppressing Diffusion-Mediated Exciton Annihilation in 2D Semiconductors Using the Dielectric Environment

Aaron J. Goodman,[1†] Der-Hsien Lien,[2†] Geun Ho Ahn,[2] Leo L. Spiegel,[3] Matin Amani,[2]

Adam P. Willard,[1] Ali Javey,[2*] William A. Tisdale[3*]

1 – Department of Chemistry, Massachusetts Institute of Technology
2 – Department of Electrical Engineering and Computer Science, University of California Berkeley
3 – Department of Chemical Engineering, Massachusetts Institute of Technology

[†] equal contribution

[*] to whom correspondence should be addressed (tisdale@mit.edu; ajavey@eecs.berkeley.edu)

Atomically thin semiconductors such as monolayer $MoS_2$ and $WS_2$ exhibit nonlinear exciton-exciton annihilation at notably low excitation densities (below ~10 excitons/µm² in $MoS_2$). Here, we show that the density threshold at which annihilation occurs can be tuned by changing the underlying substrate. When the supporting substrate is changed from $SiO_2$ to $Al_2O_3$ or $SrTiO_3$, the rate constant for second-order exciton-exciton annihilation, $k_{XX}$ [cm²/s], is reduced by one or two orders of magnitude, respectively. Using transient photoluminescence microscopy, we measure the effective room-temperature exciton diffusion coefficient in chemical-treated $MoS_2$ to be $D = 0.06 \pm 0.01$ cm²/s, corresponding to a diffusion length of $L_D = 350$ nm for an exciton lifetime of $\tau = 20$ ns, which is independent of the substrate. These results, together with numerical simulations, suggest that the effective exciton-exciton annihilation radius monotonically decreases with increasing refractive index of the underlying substrate. Exciton-exciton annihilation limits the overall efficiency of 2D semiconductor devices operating at high exciton densities; the ability to tune these interactions via the dielectric environment is an important step toward more efficient optoelectronic technologies featuring atomically thin materials.

Since the discovery[1] of monolayer and atomically thin transition metal dichalcogenides (TMDs), the diverse physics of strongly bound and highly absorbing excitons in TMD monolayers has attracted interest in these materials. The unique dielectric environment resulting from atomically-thin high index media produces anomalous Coulomb interactions[2-4] resulting in stable excitons, trions,[5-6] and biexcitons.[7] Simultaneously, atomically thin optoelectronic devices such as transistors,[8-9] phototransistors,[9] and LEDs[10] have been fabricated. Heterostructures composed of multiple TMDs[11-14] as well as TMDs paired with other complementary nanostructures[4, 15-16] have been explored.

Exciton transport and annihilation in TMDs has previously been characterized using a variety of steady state and time-resolved techniques.[17-23] Transient absorption microscopy[20, 24-28] and transient photoluminescence microscopy[29] have been particularly powerful approaches, since these techniques allow the spatial extent of the exciton population to be directly visualized.[21, 22-23] Exciton diffusivities as large as ~2 cm$^2$/s have been measured in exfoliated WSe$_2$[29] and WS$_2$.[20] However, because the as-exfoliated monolayers are inherently doped and predominately filled with other quasiparticles such as trions, the interplay between exciton diffusivity, defect states, and exciton-exciton interactions is not clear.

Recently, Javey and co-workers described a bis(trifluoromethane)sulfonamide (TFSI) treatment that substantially improves the photoluminescence quantum yield (PL QY) of TMDs.[30-32] The QY approaches unity at low excitation densities,[30] but many optoelectronic devices such as light emitting devices and lasers require much higher exciton densities. In this regard TMDs can perform poorly due to exciton-exciton annihilation, a non-radiative decay pathway that dominates at high carrier densities. The low dimensionality of monolayer TMDs as well as the strong many-body interactions characteristic of atomically thin materials leads to low PL QY at even moderate exciton densities.[30]

Here, we use time- and spatially-resolved photoluminescence spectroscopy to measure exciton transport and annihilation in TFSI-treated MoS$_2$ and WS$_2$ supported on quartz, sapphire, and strontium titanate (STO). We show that the exciton-exciton annihilation rate constant, $k_{XX}$, decreases by nearly two orders of magnitude when the substrate is changed from quartz to STO. Since the recombination is dominated by neutral excitons after TFSI treatment, we are able to directly measure the exciton diffusivity, $D$, to be 0.06 ± 0.01 cm$^2$s$^{-1}$ in MoS$_2$ using transient PL microscopy, and show that this value does not depend on the substrate. We discuss possible mechanisms behind these observations and propose a model of exciton trapping and annihilation in TMDs, where exciton-exciton annihilation proceeds primarily through interaction of freely mobile excitons with spatially localized trapped excitons. Finally, we discuss the importance of suppressing exciton annihilation for the realization of optoelectronic devices operating at high exciton densities, including high brightness LEDs, lasers, and polaritonic devices.

**Results**

MoS$_2$ monolayers were mechanically exfoliated from bulk single crystals onto SiO$_2$/Si substrates. The samples were transferred to quartz (amorphous SiO$_2$, $n$ = 1.45), sapphire (crystalline Al$_2$O$_3$, $n$ = 1.76), or strontium titanate ("STO" = crystalline SrTiO$_3$, $n$ = 2.50), substrates chosen for their varied dielectric constant. MoS$_2$ samples were then encapsulated with a poly(methyl methacrylate) (PMMA) capping layer and treated according to the TFSI treatment detailed by Amani *et al.*[30-31] The PMMA encapsulation mimics a similar strategy described in the literature by Kim *et al*. that uses fluoropolymer encapsulation to stabilize the TFSI treatment against solvent washing and vacuum exposure.[33] An optical transmission micrograph of an exemplar flake is shown in Fig. 1a and the corresponding substrate/sample/polymer stack is illustrated in Fig. 1b.

*Exciton Annihilation Rate*

After the TFSI treatment, the PL of the MoS$_2$ monolayers supported on quartz, sapphire, and STO were markedly enhanced (normalized PL spectra are shown in Fig. 1c, where the peak energies and profiles are identical; corresponding absorption spectra are shown in the supplemental information). Calibrated PL intensity was measured as a function of the generation rate, allowing the extraction of steady-state QYs, which are plotted in Fig. 1d. The QY series for MoS$_2$ monolayers supported by all three substrates are qualitatively similar; the QY was observed to be near unity at low generation rate and then decreased as the generation rate increased. The QY's dependence on the generation rate can be described by the ratio of the radiative decay rate to the sum of the rates of all decay paths available to the exciton:

$$\text{QY} = \frac{k_\text{X} N}{k_\text{X} N + k_\text{NR} N + k_\text{XX} N^2}, \tag{1}$$

where $k_\text{X}$ is the radiative decay rate, and $k_\text{NR}$ and $k_\text{XX}$ are the first order nonradiative decay rate and the exciton-exciton annihilation rate respectively. This recombination model successfully captures the PL behavior at high generation rate (i.e. high laser power) where the QY drops precipitously because exciton-exciton annihilation ($k_\text{XX} N^2$) begins to outpace radiative decay ($k_\text{X} N$). Note that the QY was near unity in all samples studied at low generation rate, showing that $k_\text{NR}$ is negligibly small and photogenerated carrier recombination is dominated by neutral excitons. This allows us to extract $k_\text{XX}$ = 0.8, 0.02, 0.005 cm$^2$s$^{-1}$ for the samples on quartz, sapphire, and STO, respectively (the dash lines in Fig 1d are the fits using Eq. 1). Notably, we found that the threshold generation rate at which the QY dropped below 50% could be increased by two orders of magnitude through changing the substrate; for samples on quartz, sapphire, and STO, those generation rates were $1.5\times10^{16}$, $2.1\times10^{17}$, and $1.1\times10^{18}$ cm$^{-2}$s$^{-1}$.

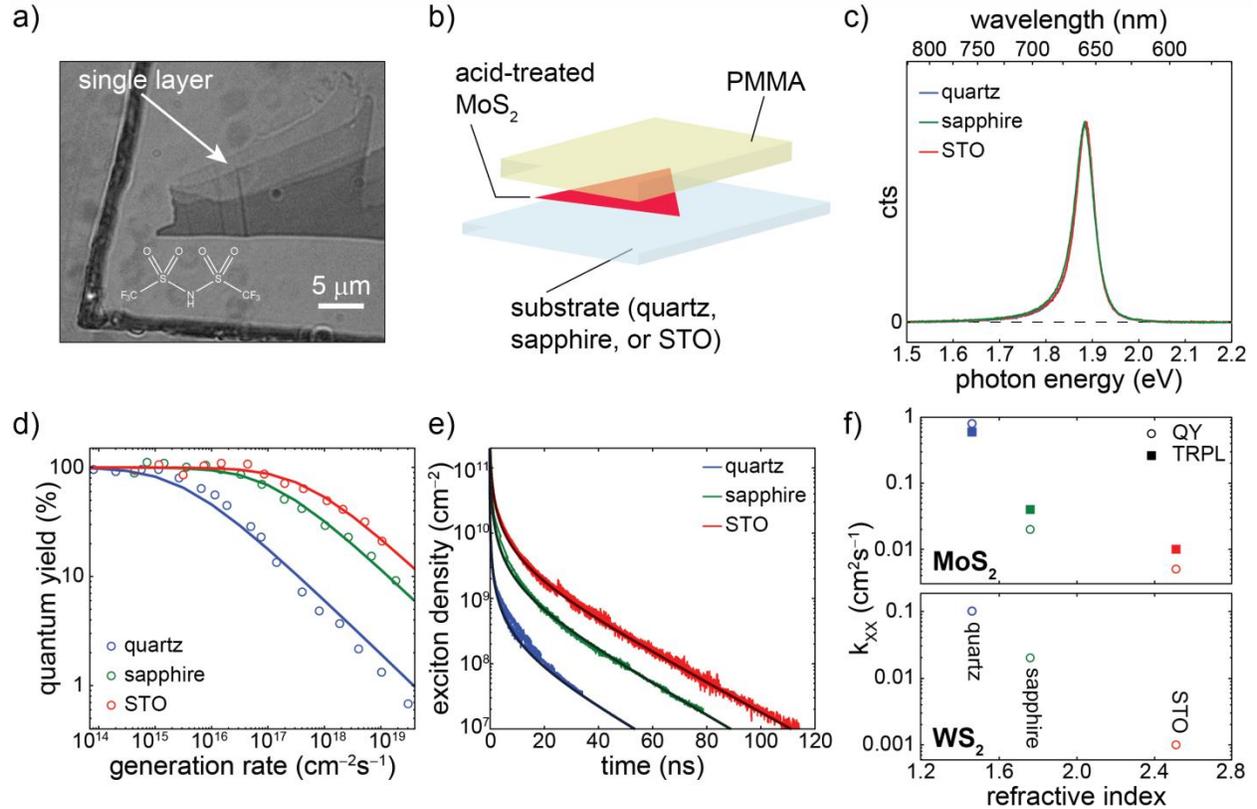

**Figure 1** *Effect of substrate on the optical properties of TFSI-treated MoS$_2$. (a) Transmission optical micrograph of exfoliated MoS$_2$ with monolayer region indicated. The inset depicts the chemical structure of the TFSI used in the treatment. The dark thick line is the border of the polymer capping layer on top of the MoS$_2$ flake. (b) Schematic of the substrate/MoS$_2$/polymer stack. (c) Photoluminescence spectra of treated MoS$_2$ on quartz (blue), sapphire (green), and STO (red). (d) Steady-state QY measured as a function of exciton generation rate. Data were recorded for MoS$_2$ supported on quartz (blue trace), sapphire (green trace), and strontium titanate (red trace). (e) Time-resolved photoluminescence traces. The traces were globally fit to extract $k_X$ and $k_{XX}$ as described in the text. (f) Exciton-exciton annihilation rate constants, $k_{XX}$, inferred from steady-state QY measurements (open circles) and time-resolved photoluminescence measurements (filled squares). $k_{XX}$ values for MoS$_2$ on quartz (blue), sapphire (green), and strontium titanate (red) are plotted against the supporting substrates' refractive indices (top). Analogous data for WS$_2$ samples are shown in the bottom panel.*

The exciton-exciton annihilation rate constant, $k_{XX}$, can also be extracted from transient measurements. Time correlated single photon counting (TCSPC) measurements were performed at varied incident laser fluences (corresponding to different initial exciton concentrations, $N(0)$) to reveal recombination dynamics. By stitching together the decay curves with varied $N(0)$, a single decay curve with over 4-decades dynamic range was obtained (individual PL decay curves are provided in the supplemental information). The decay curves for treated $MoS_2$ supported by all three substrates are shown in Fig. 1e. The decay curves are multiexponential containing fast components due to annihilation at high exciton density and slower radiative decay at low exciton density. The rates $k_X$ and $k_{XX}$ can be extracted by fitting the decay curves to a simple kinetic model, in which the excited exciton density, $N(t)$, decays according to the equation,

$$\frac{dN(t)}{dt} = -k_X N(t) - k_{XX} N^2(t). \qquad (2)$$

The values of $k_{XX}$ obtained by this fitting for TCSPC are in good agreement with the values extracted from the steady-state QY measurements, as plotted in the top panel of Fig. 1f. Notably, we observed that $k_{XX}$ varied similarly in $WS_2$ with changing supporting substrate. The values of $k_{XX}$ found in $WS_2$ are shown in the bottom panel of Fig. 1f (detail in supplemental information).

*Exciton Diffusion Imaging*

The Bohr radius of the lowest-energy band-edge exciton in $MoS_2$ has been calculated to be roughly 5-10 Å.[3, 34] In quartz-supported $MoS_2$, the photoluminescence QY dropped to 90% at a generation rate of 0.5 excitons $\mu m^{-2}$ per exciton lifetime (20 ns). One potential explanation for exciton-exciton annihilation at such a small generation rates is highly diffusive excitons that diffuse long distances before meeting and annihilating.

To probe exciton transport, we followed exciton motion in space and time using transient photoluminescence microscopy.[22, 35-36] The optical setup is depicted in Fig. 2a. A pulsed laser is focused to a diffraction-limited excitation spot at the sample using an oil-immersion objective, and the epi-fluorescence is collected by the same objective. A 360x magnified image of the fluorescing exciton population is scanned by an APD detector, which is synchronized to the pulsed laser to collect PL decay histograms. A PL decay trace was collected at each detector position in the image plane effectively recording a video of exciton transport and decay pixel-by-pixel.

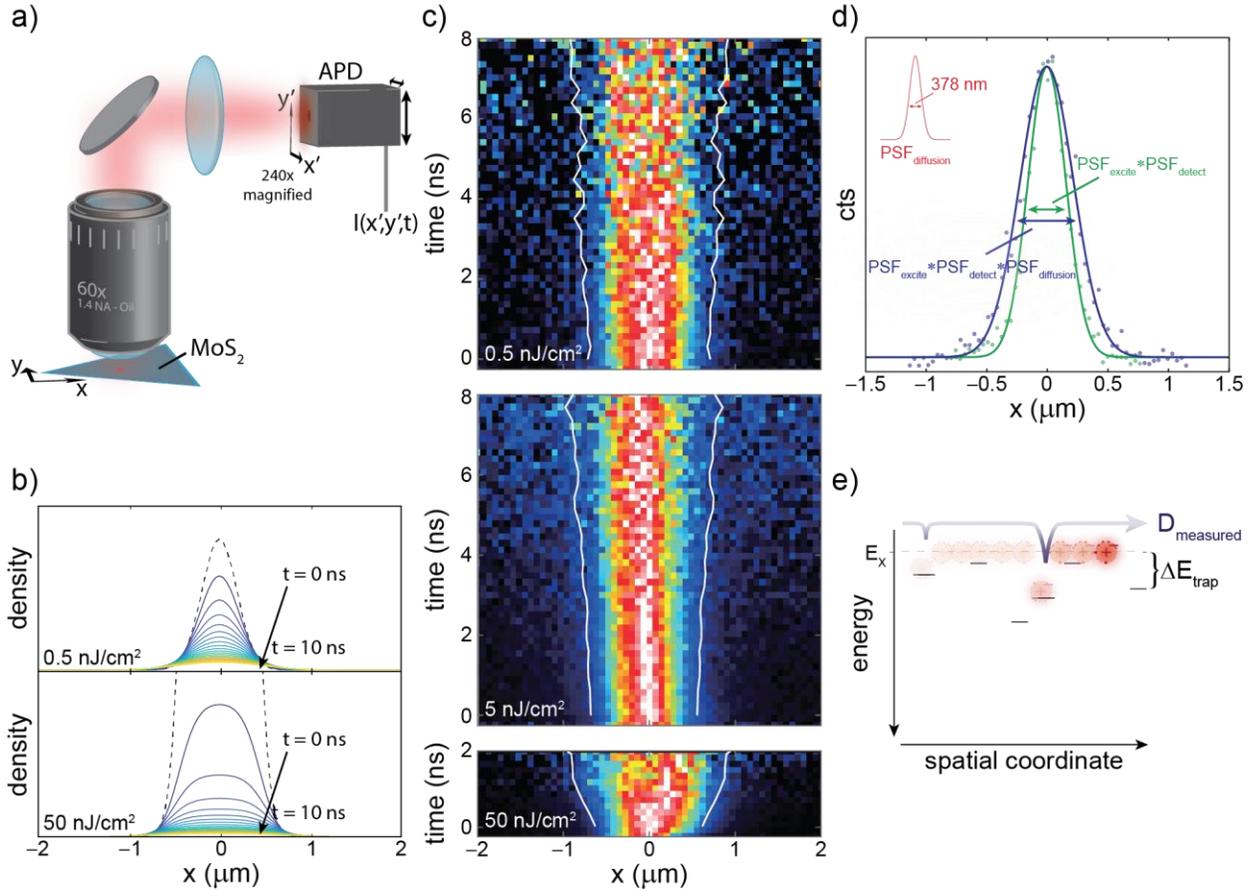

**Figure 2** *Exciton diffusion imaging. (a) Schematic diagram of the optical apparatus used to resolve exciton diffusion in space and time. (b) top: simulation of the photoluminescence intensity along a line cut of the radially symmetric exciton population as a function time. With an incident fluence of 0.5 nJ/cm$^2$, almost all excitons decay radiatively and the spot broadens due to diffusion. bottom: same as top panel, but with an incident fluence of 50 nJ/cm$^2$, many excitons decay due to annihilation. This results in artificial broadening of the density profile; excitons decay most rapidly in the center of the spot where exciton density is highest. (c) Experimentally observed broadening of the exciton population with time at 0.5 nJ/cm$^2$, 5 nJ/cm$^2$, and 50 nJ/cm$^2$ incident fluences in the top, middle, and bottom panels, respectively. White lines indicate the evolution of the standard deviation with time. (d) Steady-state measurement of exciton diffusion. The intensity profile of luminescence coming from a film of quantum dots is indicated in green. Excitons do not diffuse in this sample, so the intensity profile indicates the performance of our optical system. The photoluminescence intensity profile collected from MoS$_2$ with a generation rate of 4.8×10$^{15}$ cm$^{-2}$s$^{-1}$ is broadened due to diffusion. The diffusive contribution to the width of the intensity distribution is indicated in red in the inset. (e) This cartoon illustrates the role that trapped excitons play in influencing the measured diffusivity. Though excitons diffuse quickly at the band edge with the band edge diffusivity, they fall into immobile traps. Consequently the measured diffusivity ($D_{measured}$) is much smaller than the true band edge diffusivity.*

The top panel of Fig. 2b depicts the simulated time evolution of an exciton population initialized with a Gaussian spatial profile, indicated by the dashed black trace, designed to mimic the exciton population instantaneously excited by a 0.5 nJ/cm² laser pulse focused to a diffraction limited spot ($\lambda$ = 405 nm). At this fluence, excitons only decay radiatively. The exciton population decays exponentially as time progresses (coded in the trace colors). Simultaneously, excitons diffuse out of the initial excitation spot, broadening the distribution. We model this decay process in terms of a continuum model in which the exciton density, $N(r,t)$, evolves as a function of space and time according to

$$\frac{dN}{dt} = D\nabla^2 N - (k_X + k_{NR})N - k_{XX}N^2, \quad (3)$$

where $D$ is the exciton diffusivity. The bottom panel of Fig. 2b depicts an analogous simulation performed with a higher excitation fluence. In this regime, the exciton-exciton annihilation term in Eq. (3) becomes prominent. Excitons still undergo radiative decay and diffusion, but additionally annihilate with a rate that depends nonlinearly on the local exciton density. The nonradiative decay channel increases the overall decay rate of the exciton population and also changes the shape of the distribution. The peak of the excited distribution decays more quickly than the tails, leading to artificial flattening and broadening.

In Fig. 2c we present the time-resolved spatial broadening of the exciton population measured in quartz-supported monolayer MoS$_2$ at three different incident laser fluences. The top panel depicts data in which each laser pulse excited on average 1.2 excitons per square micron. The white traces track the standard deviation of the distribution with time. For purely diffusive broadening, the change in the distribution variance grows linearly in time:

$$\sigma^2(t) - \sigma^2(0) = 2Dt. \quad (4)$$

This behavior is observed at low fluences, and fitting the variance as a function of time allows us to extract a diffusivity $D_{qtz}$ = 0.06 ± 0.01 cm²/s, corresponding to a diffusion length $L_D = \sqrt{D\tau}$ = 350 nm. In contrast, with higher excitation fluences, the spot appears to broaden more quickly. However, this is due to faster rates of exciton-exciton annihilation in the center of the distribution rather than faster exciton transport. In these cases, the variance of the intensity distribution $I(x,t)$ grows sublinearly in time. Such details are captured by the data and reproduced by the simulation parameterized by our measured values for $k_{XX}$ and $D$ (see supplemental information). We performed the same measurement for samples supported on sapphire and STO (see supplemental information) and extracted diffusivities $D_{sapphire}$ = 0.04 ± 0.01 cm²/s and $D_{STO}$

= 0.06 ± 0.02 cm$^2$/s. The choice of substrate did not appear to significantly affect the exciton diffusivity.

To corroborate the time-resolved measurement of the diffusion coefficient on quartz, we performed a separate measurement of exciton diffusion length using steady-state PL imaging. A CW laser was focused to a diffraction limited spot at the sample and the emission was imaged on a CCD camera. To calibrate the optical system, we first performed the experiment with a thick film of well-separated CdSe quantum dots, in which no exciton diffusion could occur (see supplemental information).[36] This demonstrated that the convolved excitation and collection point spread functions (PSFs) of our microscope are nearly diffraction limited (339 nm measured *vs* 304 nm expected, see Fig. 2d). We then performed the same measurement with MoS$_2$ at sufficiently low fluence to avoid exciton-exciton annihilation. Under the same conditions, the spot is broadened due to exciton diffusion, as shown in Fig. 2d. During the excitons' lifetime, the variance increased by 0.026 μm$^2$ implying a diffusivity of 0.03 ± 0.01 cm$^2$/s, which is consistent with the time-resolved measurements.

**Discussion**

*Room-Temperature Photoluminescence*

As shown in Fig. 1c, the room-temperature photoluminescence spectrum of MoS$_2$ was unchanged when the sample was transferred from quartz to sapphire to STO, implying that neither the quasiparticle gap nor the exciton binding energy was affected by changing the dielectric constant of the supporting substrate. This observation is in contrast to theoretical predictions[34, 37-38] but consistent with experimental observations by other groups.[39-40] Moreover, we also found that the first-order decay constant, $k_X$, remained constant for MoS$_2$ on all three substrates (see low-density regime of Fig. 1e). This differs from III-V thin film devices in which the radiative recombination rate is highly dependent on the optical mode density and refractive index of the medium.[41] The observation of substrate-invariant $k_X$ in treated MoS$_2$ is consistent, however, with our previously published understanding that the measured value of $k_X$ in TFSI-treated MoS$_2$ is primarily determined by the residence time of excitons in long-lived dark states, rather than by the intrinsic strength of the band-edge optical transition.[42]

*Low-Temperature Photoluminescence*

When the sample was cooled to 77 K, differences in the photoluminescence spectrum among the three substrates were observed. Fig. 3a shows the photoluminescence spectra collected at 77 K for TFSI-treated MoS$_2$ supported on quartz, sapphire, and STO. Note that the original room-temperature PL intensity and spectrum were recovered after the sample was heated from 77 K back to room temperature (see Supplementary Information). The spectra (plotted on a logarithmic scale) exhibit a dominant peak centered at roughly 1.95 eV that is characteristic of band-edge exciton emission. The three spectra also exhibit low energy emission below the free exciton, which is not evident in room temperature photoluminescence spectra. These low energy features correspond to emission from weakly-radiative "dark" exciton states, which we previously assigned to structural defects in the native MoS$_2$ crystal.[42-43] The spectra in Fig. 3a are successfully fit to a Fermi-Dirac distribution over a Gaussian density of band-edge states plus an exponential tail density of defect states below the band edge (see Supplementary Information). While performing all three fits, shown by the red lines in Fig. 3a, the ratio of the degeneracy of the band edge and trapped exciton states was held constant (i.e. we assumed that the trap states were intrinsic to MoS$_2$ and their density did not depend on the substrate).

The energetic distribution of the trap states, parametrized by the exponential decay constant $\alpha$ [eV$^{-1}$] annotated in the figure, was obtained from fitting the PL spectrum. As the refractive index of the substrate increases, the trap state distribution becomes shallower. The sensitivity of the trap state energy to the surrounding dielectric environment and the low oscillator strength of radiative recombination from these states ($\tau_{rad,trap} \sim 10^{-6}$ s)[42] is consistent with a charge-separated exciton state exhibiting weak overlap in the electron-hole wave function. Such states may arise from trapped charges at the oxide interface. However, the monotonic dependence of the trap state distribution on the substrate dielectric constant considered together with work from other groups[43] suggests that the most likely origin of the sub-band PL shown in Fig. 3a is emission from excitons bound to sulfur site vacancies in the native MoS$_2$ crystal.

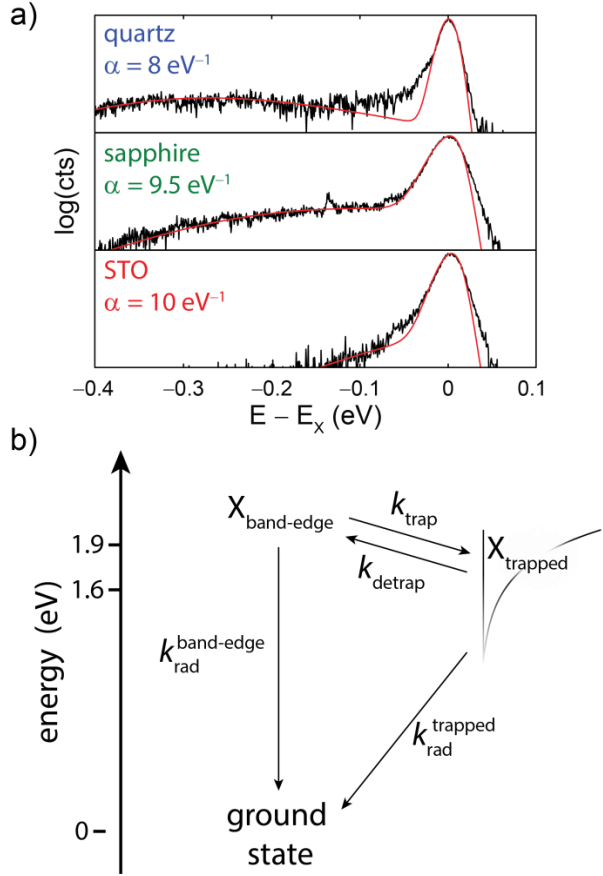

**Figure 3.** *Trapped exciton states and their equilibration with the band edge. (a) 77 K photoluminescence spectra of MoS$_2$ supported on quartz (top), sapphire (middle), and strontium titanate (bottom). At low temperature, trapped excitons emit at energies below the free exciton. The photoluminescence spectra (solid black lines) are modelled as a Fermi-Dirac distribution over an exponential density of states (red line, see supplemental information). The exponential decay constant, α [eV$^{-1}$], was extracted from the red line fit to the PL data. (b) Excitonic state diagram for trapped and band-edge (free) excitons.*

*Exciton Diffusion Coefficient*

The measured exciton diffusivity in MoS$_2$, $D \approx 0.03\text{-}0.06$ cm$^2$/s, which we obtained by two independent methods (transient imaging and CW imaging – Fig. 2 and Fig. S6), is surprisingly small. Exciton diffusivities as large as ~2 cm$^2$/s have been measured in exfoliated WSe$_2$[29] and WS$_2$[20] at higher excitation density. In a recent report, Kulig *et al.* used transient PL imaging to measure the density-dependent exciton diffusivity in freestanding and SiO$_2$-supported WS$_2$[22] and consistently obtained a value close to 0.3 cm$^2$/s in the low-density limit. Notably, this value was independent of the presence or absence of the SiO$_2$/Si substrate, in agreement with our finding that the exciton diffusivity does not depend strongly on the choice of supporting substrate.

Our observation of low exciton diffusivity at very low excitation density is again consistent with our understanding of exciton trapping in spatially localized structural defects. In previous work we showed that, at room temperature, excitons in TFSI-treated MoS$_2$ spend roughly 95% of their lifetime immobilized in dark states below the band edge.[42] Equilibration between dark trapped exciton states and bright band edge exciton states lengthens the effective exciton lifetime from ~500 ps to ~20 ns at room temperature.[35] The experimentally observed exciton diffusivity, which is a time-weighted average of the free and immobile states, is small despite the fact that excitons may diffuse very quickly while at the band edge. This principle is illustrated in Fig. 2e. By dividing the measured diffusivity ($D_\text{measured}$) by the ratio of the band edge exciton radiative rate and the apparent radiative rate ($\tau_\text{rad}$ and $\tau_\text{apparent}$ respectively) we can infer the band edge diffusivity from our measurement:

$$D_\text{band edge} = D_\text{measured} \times \left(\frac{\tau_\text{apparent}}{\tau_\text{rad}}\right). \qquad (5)$$

This leads to an inferred band edge diffusivity of $D_\text{band edge} = 2 \pm 1$ cm$^2$/s (using $\tau_\text{apparent} = 20 \pm 10$ ns), which is close to the expected diffusivity based on simple kinetic models[22] and measurements performed by other groups at higher excitation densities.[20]

*Exciton Annihilation Rate*

The simultaneous observations of low exciton diffusivity and exciton-exciton annihilation at very low exciton density suggests that excitons annihilate over distances much larger than the exciton Bohr radius, and that these exciton-exciton interactions are more effectively screened by higher-index substrates. While intriguing, such a hypothesis is difficult to reconcile with our observation of unchanging PL emission energy. How can the substrate dielectrically screen exciton-exciton annihilation (presumably a dipole-dipole interaction) without affecting the exciton binding energy or quasiparticle gap? Given the long residence time of excitons in trapped states, a more likely scenario to consider is one in which annihilation is dominated by encounters between freely mobile excitons and immobilized trapped excitons. Since the trap state distribution was experimentally observed to depend on the substrate (Fig. 3a), it is most plausible that these are the states involved in the substrate-dependent annihilation process.

To test our physical model of the substrate dependence, we performed Monte Carlo simulations of exciton transport and annihilation including both trapped and freely-mobile excitons. The results of these simulations, shown in Fig. 4, quantitatively reproduced all of our experimental observations. Key parameters involving the interaction between trapped and freely diffusing excitons are illustrated in Fig. 4a. In the model, long-lived, immobile trapped excitons act as nonradiative recombination centers for diffusing excitons. When diffusing excitons come within a critical radius $R$ of a trapped excitons, they annihilate in a bimolecular process. While many trapped excitons have the opportunity to detrap and either decay radiatively or find another trapped exciton to annihilate, a non-negligible portion of the trapped exciton population (which spans roughly 400 meV) does not possess sufficient thermal energy to detrap. This subpopulation of deeply trapped excitons persists for a long time when compared to the apparent exciton lifetime (10s of ns) and the band edge radiative lifetime (less than 1 ns).

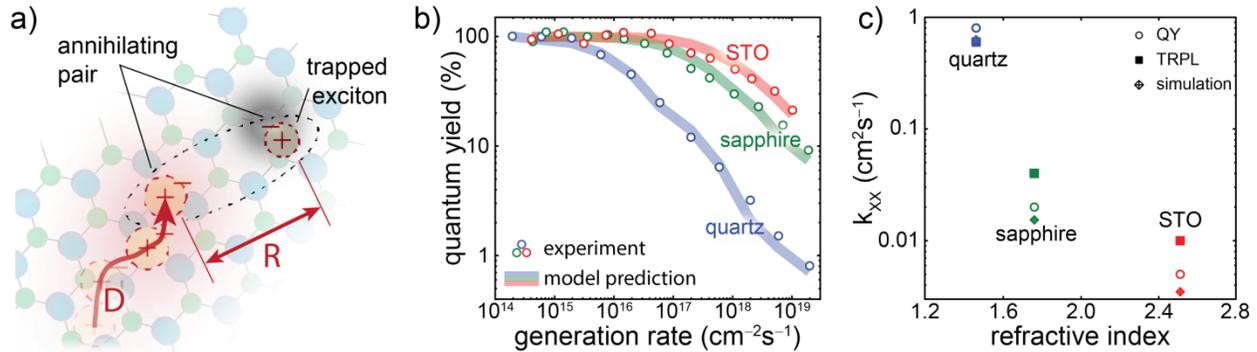

**Figure 4** *Monte Carlo simulation of trapped excitons contributing to annihilation. (a) Diffusing excitons meet long-lived trapped excitons and annihilate. The equilibrium between trapped and free excitons is responsible for the large annihilation rate constants. (c) A Monte Carlo model invoking trapped and diffusing band edge excitons quantitatively reproduces steady-state QY data. The model output is plotted here (thick transparent lines) against the experimental data (open circles) for MoS$_2$ on quartz (blue), sapphire (green), and strontium titanate (red). (c) The annihilation rate constant inferred from the model (crossed diamonds) is plotted along with the experimentally derived values (open circles, QY; filled squares, time-resolved photoluminescence). The annihilation rate is plotted against the supporting substrates' refractive indices.*

Our model describes excitons as mobile particles in a two-dimensional landscape that includes explicit spatially localized traps. Excitons were initialized at the band edge, with a spatial distribution coinciding with that of the focused laser pulse, and then allowed to diffuse in two-dimensions. The dynamics of an exciton persisted until the exciton underwent either radiative decay, simulated stochastically with a rate $k_X$, or exciton-exciton annihilation, by diffusing within a separation $R$ of another exciton. Exciton trapping and detrapping was simulated using a Monte Carlo algorithm. Namely, when an exciton encountered a trap, it trapped with unity probability and became immobilized. A trapped exciton detrapped stochastically with a rate determined by the energetic depth of the trap it occupied, as given by,

$$k_{\text{detrap}} = f_{\text{detrap}} \exp\left[-\frac{\Delta E_{\text{trap}}}{k_B T}\right], \tag{6}$$

where $f_{\text{detrap}}$ specifies the base detrapping attempt frequency, $\Delta E_{\text{trap}}$ denotes the depth of the trap, and $k_B T$ is the Boltzmann constant times temperature. Formulated in this way, the trapping and detrapping rates obey detailed balance. Traps were distributed randomly in space and each trap was assigned a random trapping energy, $\Delta E_{\text{trap}}$, distributed on the interval (0.15, 0.4) eV and weighted by exponential distribution $\rho(\Delta E_{\text{trap}}) \propto \exp[-\alpha \Delta E_{trap}]$, where the parameter $\alpha$ was obtained from fitting the low-temperature PL spectrum (Fig. 3a).

The model required very few parameters for input, most of which were available from independent experiments. The parameters and their values used in the model are enumerated in Table S1. The band edge diffusivity was modelled as $D_{\text{band edge}} = 5.0$ cm$^2$s$^{-1}$. The trapping density was chosen based on the spatial density of charged point defects observed in scanning tunneling microscopy images of exfoliated MoS$_2$,[31] which reveal a defect density of roughly 1000 per square micron. In order to reproduce the experimentally measured apparent exciton lifetime (tens of nanoseconds), it was necessary to choose a detrapping attempt frequency $f_{\text{detrap}} = 5 \times 10^{13}$ s$^{-1}$.

There are two critical parameters that depend on the supporting substrate refractive index: $\alpha$, which parameterizes the distribution of trap energies, and $R$, the radius at which two excitons annihilate. The parameter $\alpha$, which was obtained from fitting the low-temperature PL spectrum, describes the energetic distribution of the trapping sites and consequently the rate of thermally activated detrapping from those sites. The second substrate-dependent parameter is the separation, $R$, at which two excitons annihilate. This parameter is fitted by comparing simulation results to the steady-state QY measurements. The fitted values of $R$ depend strongly on the supporting substrate as well: $R = 4.0$ nm for quartz, $R = 0.6$ nm for sapphire, and $R = 0.4$ nm for STO. As the refractive index of the substrate is increased, the separation, $R$, at which excitons interact decreases. This is consistent with exciton-exciton annihilation mechanisms that require electron wavefunction overlap, or interaction *via* Coulomb or dipole interactions, both of which would be modulated by the surrounding dielectric environment, with less screening (i.e. lower index) favoring more distant interactions (i.e. larger *R*). However, we emphasize that the value of *R* may also be interpreted as an annihilation cross-section that does not have direct meaning as a physical distance of interaction.

The output of this model is plotted along with the experimental steady-state QY data in Fig. 4b; the values of $k_{XX}$ predicted by the model are plotted with the experimentally determined values in Fig. 4c. The model accurately reproduces the steady-state experiments. Importantly, the model also reproduces the annihilation dynamics and transport behavior observed in the transient experiments (see supplemental information).

**Conclusions**

Though TMD lasers[44-455] and LEDs[10, 46-477] have been demonstrated, practical use requires operation at high exciton densities. For instance, a MoTe$_2$ laser[45] exhibited a threshold pump generation rate, $R \approx 4 \times 10^{18}$ cm$^{-2}$s$^{-1}$, in the regime where exciton-exciton annihilation is dominant. Achieving high brightness LEDs or sufficiently high exciton densities for lasing or polariton condensation in the presence of competitive second order nonradiative decay channels necessitates excessive pump rates. Exciton-exciton annihilation places a fundamental limit on the operating efficiency of such devices. Understanding the mechanism behind this efficiency loss and raising the maximum achievable operating efficiency by tuning the dielectric environment are critical advances for the future of TMD optoelectronic devices.

Strong exciton-exciton interactions and the tuning of those interactions through the surrounding dielectric are both manifestations of reduced dielectric screening in 2D materials. Coulomb interactions are poorly screened in monolayer TMDs[2, 15, 40, 488] resulting in large exciton, trion, and biexciton binding energies. Though these many-body interactions can be exploited to observe physics unique to 2D materials, here they facilitate exciton-exciton annihilation, limiting radiative efficiency. Dielectric screening is weak in these materials because coulomb interactions can circumvent the highly polarizable TMD by leaking through the surrounding environment. We take advantage of that fact here by tuning the dielectric constant in the external environment to suppress exciton-exciton interactions. Tuning the strength of many-body interactions through the dielectric environment is a powerful design paradigm unique to low-dimensional materials.

**Methods**

*Sample Preparation* – Mineral $MoS_2$ (SPI) was exfoliated on $SiO_2$/Si substrates and then transferred to other substrates including quartz, sapphire or STO substrates by a dry transfer technique *via* a polymethyl methacrylate (PMMA) membrane as a transfer media. The transferred $MoS_2$ was treated by the following procedure: 20 mg of bis(trifluoromethane)sulfonamide (TFSI) was dissolved in 5 ml of 1,2-dichloroethane and then diluted with 45 ml 1,2-dichlorobenzene. The transferred $MoS_2$ with PMMA was then immersed in the TFSI solution for 30 seconds in room temperature. The sample was blow dried with nitrogen. Note that the enhancement is depending on the initial quality of the sample and only a portion of sample can reach > 95% QY.

*Transient PL* – Samples were excited using a 405 nm pulsed laser diode (Picoquant, LDH-D-C-405M, 40 MHz repetition rate, 0.4 ns pulse duration) with fluences as indicated in the main text. The laser was focused to a nearly diffraction-limited spot (Nikon, CFI S Plan Fluor ELWD, 40×, 0.6 NA). Fluorescence was collected with the same objective, and passed through a dichroic mirror and 600 nm - 700 nm bandpass filter before being focused onto a Si avalanche photodiode (Micro Photon Devices, PDM50, 50 ps resolution at the detection wavelength). The detector was connected to a counting board for TCSPC (Picoquant, PicoHarp 300).

*Quantum Yield Measurement* – The calibrated PL QY measurement has been previously described in detail.[30] Briefly, the 514.5 nm line of an Ar ion laser (Lexel 95) was focused to the sample using a 60× ultra-long working distance objective (NA = 0.7). PL was collected by the same objective, filtered and dispersed by a spectrograph. The emission was detected by a Si CCD camera (Andor, iDus BEX2DD). The excitation power and optical system spectral sensitivity were externally calibrated. The instrument function was cross-calibrated using rhodamine 6G (QY close to 100%) and spectralon as reference samples. The measured PL spectra were integrated and converted into external quantum efficiencies and corresponding QYs.

*Transient PL Microscopy* – $\lambda = 570$ nm pulses from a synchronously pumped optical parametric oscillator (Coherent, PP automatic, 76 MHz, < 1 ps) were spatially filtered by a single-mode optical fiber and used to excite the sample. The laser was focused to a diffraction-limited spot (Nikon, CFI Plan Apo Lambda, 60× Oil, 1.4 NA). Fluorescence was collected by the same objective and filtered by a dichroic mirror and 600 nm – 700 nm bandpass filter. The APD detector was placed in the 360× magnified image plane outside the microscope. The detector position in the image plane was controlled by two orthogonal motorized actuators (Thorlabs, ZFS25B). The evolution of the photoluminescence spatial profile with time was acquired by scanning the detector across the magnified emission profile and collecting a photoluminescence decay histogram at each position.

*PL Spectroscopy, Low Temperature* – A 532 nm continuous wave laser (Coherent, Sapphire SF 532-20 CW) was focused at the sample (Nikon, CFI S Plan Fluor ELWD, 40×, 0.6 NA). Fluorescence was collected by the same objective and filtered by a dichroic mirror before being dispersed by a spectrograph (Princeton Instruments, Acton SP2500) and imaged on a cooled CCD camera (Princeton Instruments, Pixis PIX100BR). Low-temperature data were collected under vacuum in a microscope-mounted cryostat (Janis, ST-500-P).

*Numerical Simulation* – Exciton dynamics were simulated with a fixed time step Monte Carlo algorithm. Excitons were initialized to the band edge according to a spatial profile matching the excitation laser intensity profile. At each time step free excitons hopped a fixed distance in a random direction. Excitons trapped with unit probability if the center position of the exciton was within 0.4 nm of the center of an empty trap. Upon moving within 2R nm of an occupied trap, the exciton annihilated and was removed from the simulation. Annihilation between pairs of band-edge excitons were rare due to the low population of detrapped excitons and were thus neglected. Trapped excitons detrapped probabilistically as described in the main text.

**Acknowledgements –** Work at MIT was supported as part of the Center for Excitonics, an Energy Frontier Research Center funded by the US Department of Energy, Office of Science, Basic Energy Sciences (BES) under Award No. DE-SC0001088 (MIT). Work at U.C. Berkeley was supported by the Electronic Materials Program funded by the Director, Office of Science, Office of Basic Energy Sciences, Materials Sciences and Engineering Division of the U.S.

Department of Energy, under contract no. DE-AC02-05Ch11231. A.J.G. acknowledges partial support from the US National Science Foundation Graduate Research Fellowship Program under Grant No. 1122374.

**Author Contributions** – A.J.G. performed time-resolved spectroscopy, diffusion imaging, and low-temperature spectroscopy under the supervision of W.A.T. D.-H.L. prepared samples and performed power-dependent quantum yield and TCSPC experiments under the supervision of A.J. A.J.G. developed and implemented the numerical model, with assistance from L.L.S., A.P.W., and W.A.T. G.H.A. assisted with sample preparation and M.A. contributed to data interpretation. All authors discussed the results and interpretation. A.J.G. and D.-H.L. wrote the manuscript with contributions from the other authors.

**Additional Information** – Supplementary information is available in the online version of the paper. Correspondence and requests for materials should be addressed to W.A.T. (tisdale@mit.edu) and A.J (ajavey@berkeley.edu).

**Competing Financial Interests** – The authors declare no competing financial interests.

**References**

(1) Mak, K. F.; Lee, C.; Hone, J.; Shan, J.; Heinz, T. F. "Atomically Thin $MoS_2$: A New Direct-Gap Semiconductor" *Phys. Rev. Lett.* **2010**, *105*, 136805.

(2) Chernikov, A.; Berkelbach, T. C.; Hill, H. M.; Rigosi, A.; Li, Y.; Aslan, O. B.; Reichman, D. R.; Hybertsen, M. S.; Heinz, T. F. "Exciton Binding Energy and Nonhydrogenic Rydberg Series in Monolayer $WS_2$" *Phys. Rev. Lett.* **2014**, *113*, 076802.

(3) Mayers, Z. M.; Berkelbach, T. C.; Hybertsen, M. S.; Reichman, D. R. "Binding Energies and Spatial Structures of Small Carrier Complexes in Monolayer Transition-Metal Dichalcogenides via Diffusion Monte Carlo" *Phys. Rev. B* **2015**, *92*, 161404.

(4) Raja, A.; Montoya-Castillo, A.; Zultak, J.; Zhang, X.-X.; Ye, Z.; Roquelet, C.; Chenet, D. A.; van der Zande, A. M.; Huang, P.; Jockusch, S.; Hone, J. C.; Reichman, D. R.; Brus, L. E.; Heinz, T. F. "Energy Transfer from Quantum Dots to Graphene and $MoS_2$: The Role of Absorption and Screening in Two-Dimensional Materials" *Nano Lett.* **2016**, (16), 2328-2333.

(5) Mak, K. F.; He, K.; Lee, C.; Lee, G. H.; Hone, J.; Heinz, T. F.; Shan, J. "Tightly Bound Trions in Monolayer $MoS_2$" *Nat. Mater.* **2013**, *12*, 207.

(6) Zhang, C.; Wang, H.; Chan, W.; Manolatou, C.; Rana, F. "Absorption of Light by Excitons and Trions in Monolayers of Metal Dichalcogenide $MoS_2$: Experiments and Theory" *Phys. Rev. B* **2014**, *89*.

(7) You, Y.; Zhang, X.-X.; Berkelbach, T. C.; Hybertsen, M. S.; Reichman, D. R.; Heinz, T. F. "Observation of Biexcitons in Monolayer WSe$_2$" *Nat. Phys.* **2015,** *11*, 477-481.

(8) Radisavljevic, B.; Radenovic, A.; Brivio, J.; Giacometti, V.; Kis, A. "Single-Layer MoS$_2$ Transistors" *Nat. Nanotechnol.* **2011,** *6*, 147-150.

(9) Baugher, B. W. H.; Churchill, H. O. H.; Yang, Y.; Jarillo-Herrero, P. "Optoelectronic Devices Based on Electrically Tunable p-n Diodes in a Monolayer Dichalcogenide" *Nat. Nanotechnol.* **2014,** *9*, 262.

(10) Ross, J. S.; Klement, P.; Jones, A. M.; Ghimire, N. J.; Yan, J.; Mandrus, D. G.; Taniguchi, T.; Watanabe, K.; Kitamura, K.; Yao, W.; Cobden, D. H.; Xu, X. "Electrically Tunable Excitonic Light-Emitting Diodes Based on Monolayer WSe$_2$ p-n Junctions" *Nat. Nanotechnol.* **2014,** *9*, 268.

(11) Fang, H.; Battaglia, C.; Carraro, C.; Nemsak, S.; Ozdol, B.; Kang, J. S.; Bechtel, H. A.; Desai, S. B.; Kronast, F.; Unal, A. A.; Conti, G.; Conlon, C.; Palsson, G. K.; Martin, M. C.; Minor, A. M.; Fadley, C. S.; Yablonovitch, E.; Maboudian, R.; Javey, A. "Strong Interlayer Coupling in van der Waals Heterostructures Built from Single-Layer Chalcogenides" *PNAS* **2014,** *111*, 6198-6202.

(12) Hong, X.; Kim, J.; Shi, S.-F.; Zhang, Y.; Jin, C.; Sun, Y.; Tongay, S.; Wu, J.; Zhang, Y.; Wang, F. "Ultrafast Charge Transfer in Atomically Thin MoS$_2$/WS$_2$ Heterostructures" *Nat. Nanotechnol.* **2014,** *9*, 682-686.

(13) Gong, Y.; Lin, J.; Wang, X.; Shi, G.; Lei, S.; Lin, Z.; Zou, X.; Ye, G.; Vajtai, R.; Yakobson, B. I.; Terrones, H.; Terrones, M.; Tay, B. K.; Lou, J.; Pantelides, S. T.; Liu, Z.; Zhou, W.; Ajayan, P. M. "Vertincal and in-Plane Heterostructures from WS$_2$/MoS$_2$ Monolayers" *Nat. Mater.* **2014,** *13*, 1135-1142.

(14) Ma, Q.; Andersen, T. I.; Nair, N. L.; Gabor, N. M.; Massicotte, M.; Lui, C. H.; Young, A. F.; Fang, W.; Watanabe, K.; Taniguchi, T.; Kong, J.; Gedik, N.; Koppens, F. H. L.; Jarillo-Herrero, P. "Tuning Ultrafast Electron Thermalization Pathways in a van der Waals Heterostructure" *Nat. Phys.* **2016,** *12*, 455-459.

(15) Prins, F.; Goodman, A. J.; Tisdale, W. A. "Reduced Dielectric Screening and Enhanced Energy Transfer in Single- and Few-Layer MoS$_2$" *Nano Lett.* **2014,** *14*.

(16) Prasai, D.; Klots, A. R.; Newaz, A.; Niezgoda, J. S.; Orfield, N. J.; Escobar, C. A.; Wynn, A.; Efimov, A.; Jennings, G. K.; Rosenthal, S. J.; Bolotin, K. I. "Electrical Control of near-Field Energy Transfer between Quantum Dots and Two-Dimensional Semiconductors" *Nano Lett.* **2015**.

(17) Shi, H.; Yan, R.; Bertolazzi, S.; Brivio, J.; Gao, B.; Kis, A.; Jena, D.; Xing, H. G.; Huang, L. "Exciton Dynamics in Suspended Monolayer and Few-Layer MoS$_2$ 2D Crystals" *ACS Nano* **2013,** *7* (2).

(18) Yuan, L.; Huang, L. "Exciton Dynamics and Annihilation in WS$_2$ 2D Semiconductors" *Nanoscale* **2015,** *7*, 7402-7408.

(19) Sun, D.; Rao, Y.; Reider, G. A.; Chen, G.; You, Y.; Brezin, L.; Harutyunyan, A. R.; Heinz, T. F. "Observation of Rapid Exciton-Exciton Annihilation in Monolayer Molybdenum Disulfide" *Nano Lett.* **2014,** *14*, 5625-5629.

(20) Yuan, L.; Wang, T.; Zhu, T.; Zhou, M.; Huang, L. "Exciton Dynamics, Transport, and Annihilation in Atomically Thin Two-Dimensional Semiconductors" *J. Phys. Chem. Lett.* **2017,** *8* (14), 3371-3379.

(21) Kato, T.; Kaneko, T. "Transport Dynamics of Neutral Excitons and Trions in Monolayer $WS_2$" *ACS Nano* **2016,** *10* (10), 9687-9694.

(22) Kulig, M.; Zipfel, J.; Nagler, P.; Blanter, S.; Schuller, C.; Korn, T.; Paradiso, N.; Glazov, M. M.; Chernikov, A. "Exciton Diffusion and Halo Effects in Monolayer Semiconductors" *Phys. Rev. Lett.* **2018,** *120*, 207401.

(23) Onga, M.; Zhang, Y.; Ideue, T.; Iwasa, Y. "Exciton Hall Effect in Monolayer $MoS_2$" *Nat. Mater.* **2017,** *16*, 1193-1197.

(24) Kumar, N.; He, J.; He, D.; Wang, Y.; Zhao, H. "Charge Carrier Dynamics in Bulk $MoS_2$ Crystal Studied by Transient Absorption Microscopy" *J. Appl. Phys.* **2013,** *113*.

(25) Kumar, N.; Cui, Q.; Ceballos, F.; He, D.; Wang, Y.; Zhao, H. "Exciton Diffusion in Monolayer and Bulk $MoSe_2$" *Nanoscale* **2014,** *6*.

(26) Cui, Q.; Ceballos, F.; Kumar, N.; Zhao, H. "Transient Absorption Microscopy of Monolayer and Bulk $WSe_2$" *ACS Nano* **2014,** *8*.

(27) Zhu, T.; Yuan, L.; Zhao, Y.; Zhou, M.; Wan, Y.; Mei, J.; Huang, L. "Highly Mobile Charge-Transfer Excitons in Two-Dimensional {$WS_2$}/Tetracene Heterostructures" *Sci. Adv.* **2018,** *4* (1).

(28) Wang, R.; Ruzicka, B. A.; Kumar, N.; Bellus, M. Z.; Chiu, H.-Y.; Zhao, H. "Ultrafast and Spatially Resolved Studies of Charge Carriers in Atomically Thin Molybdenum Disulfide" *Phys. Rev. B* **2012,** *86*.

(29) Mouri, S.; Miyauchi, Y.; Toh, M.; Zhao, W.; Eda, G.; Matsuda, K. "Nonlinear Photoluminescence in Atomically Thin Layered $WSe_2$ Arising from Diffusion-Assisted Exciton-Exciton Annihilation" *Phys. Rev. B* **2014,** *90*.

(30) Amani, M.; Lien, D.-H.; Kiriya, D.; Xiao, J.; Azcatl, A.; Noh, J.; Madhvapathy, S. R.; Addou, R.; KC, S.; Dubey, M.; Cho, K.; Wallace, R. M.; Lee, S.-C.; He, J.-H.; Ager III, J. W.; Zhang, X.; Yablonovitch, E.; Javey, A. "Near-Unity Photoluminescence Quantum Yield in $MoS_2$" *Science* **2015,** *350* (6264).

(31) Amani, M.; Taheri, P.; Addou, R.; Ahn, G. H.; Kiriya, D.; Lien, D.-H.; Ager III, J. W.; Wallace, R. M.; Javey, A. "Recombination Kinetics and Effects of Superacid Treatment in Sulfur- and Selenium-Based Transition Metal Dichalcogenides" *Nano Lett.* **2016,** *16*, 2786-2791.

(32) Amani, M.; Burke, R. A.; Ji, X.; Zhao, P.; Lien, D.-H.; Taheri, P.; Ahn, G. H.; Kirya, D.; Ager III, J. W.; Yablonovitch, E.; Kong, J.; Dubey, M.; Javey, A. "High Luminescence Efficiency in $MoS_2$ Grown by Chemical Vapor Deposition" *ACS Nano* **2016,** *10*, 6535-6541.

(33) Kim, H.; Lien, D.-H.; Amani, M.; Ager, J. W.; Javey, A. "Highly Stable Near-Unity Photoluminescence Yield in Monolayer $MoS_2$ by Fluoropolymer Encapsulation and Superacid Treatment" *ACS Nano* **2017,** *11*, 5179.

(34) Ugeda, M. M.; Bradley, A. J.; Shi, S.-F.; da Jornada, F. H.; Zhang, Y.; Qiu, D. Y.; Ruan, W.; Mo, S.-K.; Hussain, Z.; Shen, Z.-X.; Wang, F.; Louie, S. G.; Crommie, M. F. "Giant Bandgap Renormalization and Excitonic Effects in a Monolayer Transition Metal Dichalcogenide Semiconductor" *Nat. Mater.* **2014**, *13*, 1091-1095.

(35) Akselrod, G. M.; Deotare, P. B.; Thompson, N. J.; Lee, J.; Tisdale, W. A.; Baldo, M. A.; Menon, V. M.; Bulovic, V. "Visualization of Exciton Transport in Ordered and Disordered Molecular Solids" *Nat. Commun.* **2014**, *5*.

(36) Akselrod, G. M.; Prins, F.; Poulikakos, L. V.; Lee, E. M. Y.; Weidman, M. C.; Mork, A. J.; Willard, A. P.; Bulovic, V.; Tisdale, W. A. "Subdiffusive Exciton Transport in Quantum Dot Solids" *Nano Lett.* **2014**, *14*.

(37) Thygesen, K. S. "Calculating Excitons, Plasmons, and Quasiparticles in 2D Materials and van der Waals Heterostructures" *2D Materials* **2017**, *4*, 022004.

(38) Qiu, D. Y.; da Jornada, F. H.; Louie, S. G. "Screening and Many-Body Effects in Two-Dimensional Crystals: Monolayer $MoS_2$" *Phys. Rev. B* **2016**, *93*, 235435.

(39) Li, Y.; Chernikov, A.; Zhang, X.; Rigosi, A.; Hill, H. M.; Van der Zande, A. M.; Chenet, D. A.; Shih, E.-M.; Hone, J.; Heinz, T. F. "Measurement of the Dielectric Funtion of Monolayer Transition-Metal Dichalcogenides $MoS_2$, MoSe2, $WS_2$, and $WSe_2$" *Phys. Rev. B* **2014**, *90*.

(40) Lin, Y.; Ling, X.; Yu, L.; Huang, S.; Hsu, A. L.; Lee, Y.-H.; Kong, J.; Dresselhaus, M. S.; Palacios, T. "Dielectric Screening of Excitons and Trions in Single-Layer $MoS_2$" *Nano Lett.* **2014**, *14*, 5569-5576.

(41) Yablonovitch, E.; Gmitter, T. J.; Bhat, R. "Inhibited and Enhanced Spontaneous Emission form Optically Thin AlGaAs/GaAs Double Heterostructures" *Phys. Rev. Lett.* **1988**, *61*, 2546-2549.

(42) Goodman, A. J.; Willard, A. P.; Tisdale, W. A. "Exciton Trapping Is Responsible for The Long Apparent Lifetime in Acid-Treated $MoS_2$" *Phys. Rev. B* **2017**, *96*, 121404(R).

(43) Atallah, T. L.; Wang, J.; Bosch, M.; Seo, D.; Burke, R. A.; Moneer, O.; Zhu, J.; Theibault, M.; Brus, L. E.; Hone, J.; Zhu, X.-Y. "Electrostatic Screening of Charged Defects in Monolayer $MoS_2$" *J. Phys. Chem. Lett.* **2017**, *8*.

(44) Wu, S.; Buckley, S.; Schaibley, J. R.; Feng, L.; Yan, J.; Mandrus, D. G.; Hatami, F.; Yao, W.; Vuckovic, J.; Majumdar, A.; Xu, X. "Monolayer semiconductor nanocavity lasers with ultralow thresholds" *Nature* **2015**.

(45) Li, Y.; Zhang, J.; Huang, D.; Sun, H.; Fan, F.; Feng, J.; Wang, Z.; Ning, C. Z. "Room-Temperature Continuous-Wave Lasing from Monolayer Molybdenum Ditelluride Integrated with a Silicon Nanobeam Cavity" *Nat. Nanotechnol.* **2017**, *12*, 987-992.

(46) Sundaram, R. S.; Engel, M.; Lombardo, A.; Krupke, R.; Ferrari, A. C.; Avouris, P.; Steiner, M. "Electroluminescence in Single Layer $MoS_2$" *Nano Lett.* **2013**, *13*, 1416-1421.

(47) Ye, Y.; Ye, Z.; Gharghi, M.; Zhu, H.; Zhao, M.; Wang, Y.; Yin, X.; Zhang, X. "Exciton-Dominant Electroluminescence from a Diode of Monolayer $MoS_2$" *Appl. Phys. Lett.* **2014**, *104*.

(48) Hill, H. M.; Rigosi, A. F.; Roquelet, C.; Chernikov, A.; Berkelbach, T. C.; Reichman, D. R.; Hybertsen, M. S.; Brus, L. E.; Heinz, T. F. "Observation of Excitonic Rydberg States in Monolayer $MoS_2$ and $WS_2$ by Photoluminescence Excitation Spectroscopy" *Nano Lett.* **2015,** *15*, 2992-2997.

Supplementary Information

for

# Suppressing Diffusion-Mediated Exciton Annihilation in 2D Semiconductors Using the Dielectric Environment

Aaron J. Goodman,[1][†] Der-Hsien Lien,[2][†] Geun Ho Ahn,[2] Leo L. Spiegel,[3] Matin Amani,[2] Adam P. Willard,[1] Ali Javey,[2][*] William A. Tisdale[3][*]

**Table of Contents:**

## I. MoS₂ Absorption Spectra

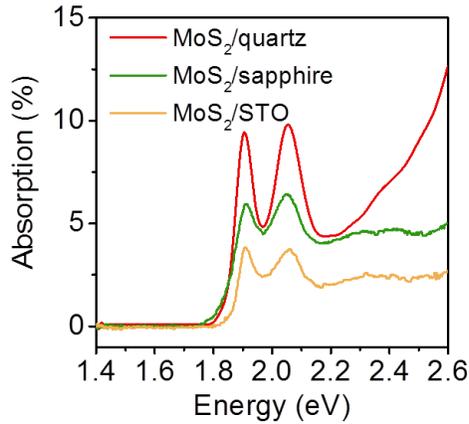

**Figure S1** *Absorption spectra of MoS₂ supported on quartz (red), sapphire (green), and STO (yellow).*

Two measurements were performed to collect the absorption spectra for 2D monolayers on quartz, sapphire and SrTiO$_3$ substrates. In the first measurement, the absolute values of reflection and transmission at a wavelength of 514.5 nm (Ar ion laser) were measured for the extraction of the absorption value using lock-in amplifier coupled with a Silicon photodiode. The extracted absorption values are 7.1%, 4.7% and 2.5% for MoS$_2$ on quartz, sapphire and STO, respectively. Absorption spectra were obtained using a supercontinuum laser (Fianium WhiteLase SC-400). The reflection/transmission light was analyzed by a spectrometer to produce the spectra. In those measurements, the laser was focused by a 50× objective. For reflection measurement, the reflected light was collected through the same objective. For transmission measurement, the transmitted light was collected using a 20× objective. The system was calibrated using quartz/sapphire/SrTiO$_3$ substrates and silver mirror as the reference for transmission and reflection. Note that the quantum yield, generation rates and exciton concentration (TRPL) are calculated by considering the absorption.

## II. PL Decay Traces

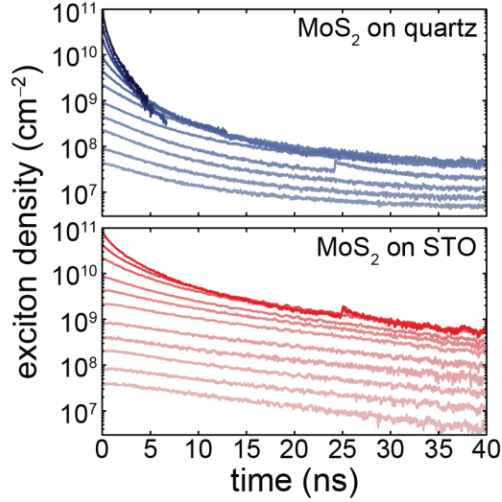

**Figure S2** *Individual PL decay histograms collected from MoS$_2$ supported on quartz (top) and STO (bottom). These decay histograms were stitched together and fit with Equation 2 in the main text to extract $k_X$ and $k_{XX}$.*

The individual PL decay histograms exhibit exciton-exciton annihilation at high fluences, which can be seen as a fast component in the PL decay trace. The multi exponential curves are fit by a continuum model in the main text according to Equation 2, reproduced here,

$$\frac{dN(t)}{dt} = -k_X N(t) - k_{XX} N^2(t). \tag{S1}$$

Such an analysis, as well as stitching together individual traces assumes that the exciton population is "well mixed", that is to say that excitons are randomly distributed throughout the course of the experiment with constant density in the probed focal volume. While this assumption holds to the extent that it is possible to extract key parameters, $k_X$, *and* $k_{XX}$, and the curves are well fit by Equation 2, one might not expect the assumption to be strictly true, particularly at high exciton densities. In particular, in the presence of annihilation, surviving excitons develop spatial correlations such that their spatial distribution isn't uniformly distributed and random. They are effectively "anti-bunched". Furthermore, the exciton density is not constant across the experimental excitation profile. However, these details do not affect the fitting of the TCSPC data as confirmed through discrete modelling that showed the continuum modelling holds up under the experimental conditions used here.

## III. WS₂ Generation Rate-Dependent Quantum Yield

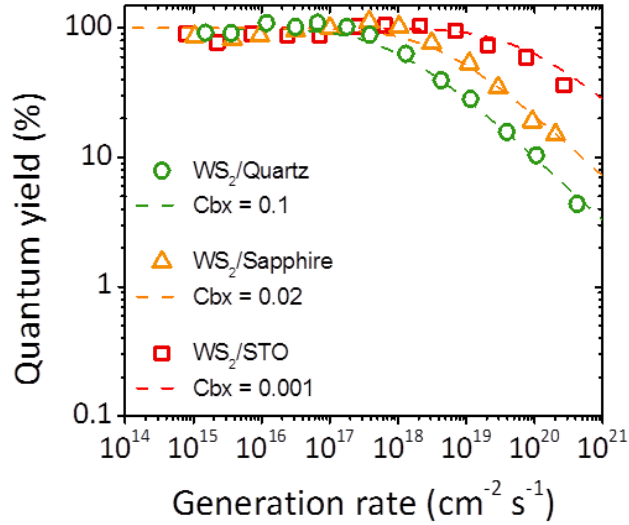

**Figure S3** *Absolute quantum yield values for WS$_2$ supported on quartz (green circles), sapphire (yellow triangles), and STO (red squares) extracted from calibrated PL spectra as a function of carrier generation rate. The data are consistent with and qualitatively similar to the data for MoS$_2$ on the three substrates. At low generation rates, the samples emit with near unity quantum yield. As the generation rate increases, the exciton-exciton annihilation interaction lowers the quantum yield dramatically.*

## IV. Continuum Model Diffusion Equation

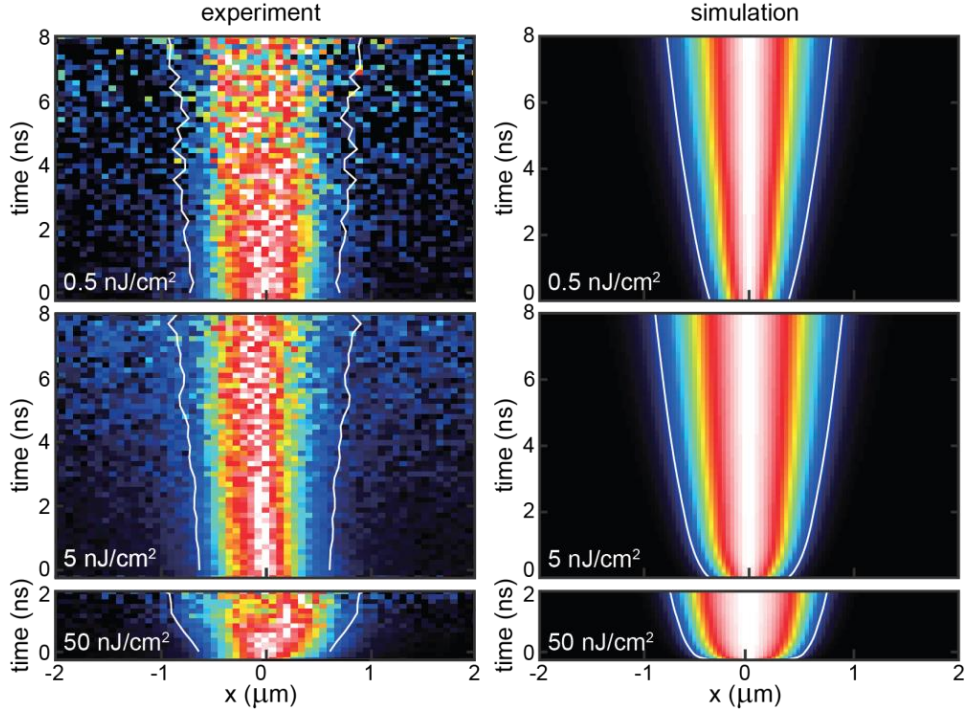

**Figure S4** *Exciton diffusivity experimental results (left) and continuum modelling (right). Horizontal cuts through the color plots represent normalized line cuts through the Gaussian exciton population spatial distribution as a function of time. White lines track $2\sigma$ where $\sigma$ is the standard deviation of the exciton spatial distribution.*

As discussed in the main text, diffusion imaging allows us to visualize the exciton population broadening in time. The experimental results reproduced in the left panels of **Fig. S4** show that the distribution broadens in time due to exciton diffusion and exciton-exciton annihilation. The experimental results are captured by a continuum model that propagates an initial Gaussian exciton population according to

$$\frac{dN}{dt} = D\nabla^2 N - (k_\text{X} + k_\text{NR})N - k_\text{XX}N^2,$$

where $D$ is the exciton diffusivity, $k_\text{XX}$ is the exciton-exciton annihilation rate constant, and $k_\text{X}$ (the apparent radiative decay rate) and $k_\text{NR}$ (the nonradiative decay rate) are first order decay rates. The model output (shown in the right panels of **Fig. S4**) agrees with experiment when parameterized with experimentally determined $D$ and $k_\text{XX}$.

## V. Diffusivity Measurements: Sapphire and STO

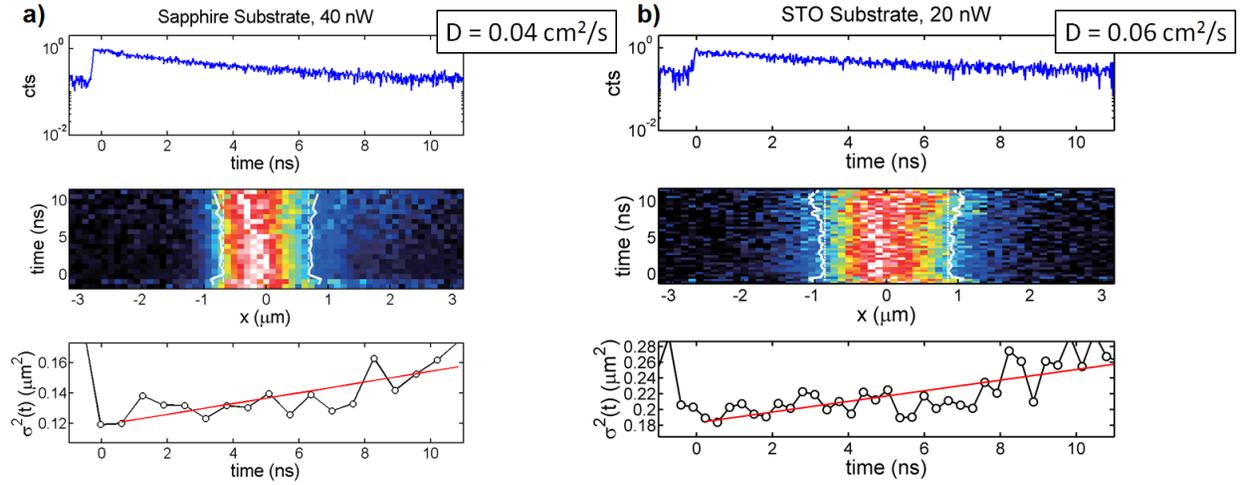

**Figure S5** *Diffusion imaging experiments performed on MoS$_2$ supported on sapphire (a), and STO (b). Top panels show the spatially-integrated PL decay. Middle panels show exciton diffusion imaging as discussed in the main text. Bottom panels show the PL spatial distribution variance extracted from the middle panels evolving in time showing diffusive broadening.*

Exciton diffusivity measurements were also performed on TFSI-treated MoS$_2$ supported on sapphire and STO. In each case the experiment was performed as described in the methods section of the main text at sufficiently low fluence to avoid exciton/exciton annihilation as indicated by the monoexponential PL decay curves shown in the top panels. The PL spatial distribution was fitted to a Gaussian at each time point. The variances from the Gaussian fits are plotted as a function of time in the bottom panel. The variance grows linearly in time at a rate that reflects the diffusivity. In both cases, the measured diffusivity was similar to the diffusivity measured for MoS$_2$ on quartz: $0.04 \pm 0.01$ and $0.06 \pm 0.02$ cm$^2$/s for MoS$_2$ on sapphire and STO respectively.

## VI. Optical Imaging System Characterization

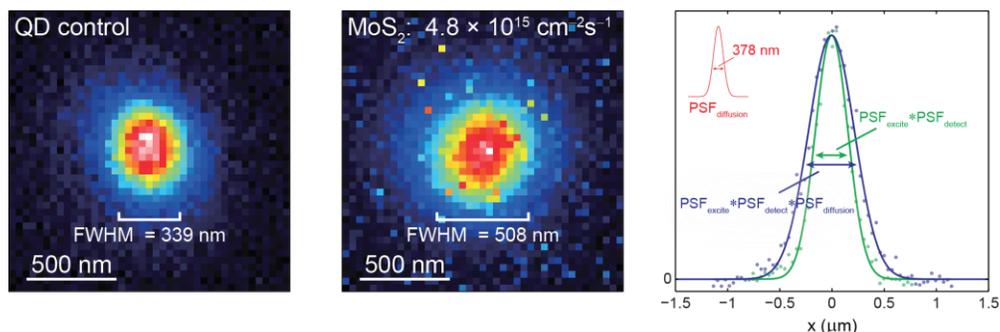

**Figure S6** *Optical imaging apparatus characterization. Left: PL image collected from a thin film of well insulated CdSe quantum dots in which exciton diffusion doesn't occur. The dots were excited with a focused laser ($\lambda = 520$ nm, NA = 1.4). Middle: PL image collected from MoS$_2$ with the same imaging system using a sufficiently low fluence to avoid exciton-exciton annihilation. Right: Radial intensity profiles of the images in the left and center panels showing that the MoS$_2$ PL intensity profile is broadened due to exciton diffusion.*

For CW imaging, the emission spot is a convolution of the excitation point spread function (PSF) (*i.e.* the Gaussian laser spot size) and the collection PSF. A diffraction limited intensity distribution of excitons is excited, and then as each emits, it is localized within the accuracy of the imaging optics. It's feasible to assess these two PSFs independently. To assess the collection PSF, a sparse film of isolated CdSe quantum dots was drop cast. This was then illuminated by an LED and the PL was imaged. Each point-like emitter appeared as a Gaussian spot with a width representing the imaging optics collection PSF.

For the excitation PSF, a complete film of dots was cast and excited with a laser. For these experiments, we used CdSe QDs coated with a thick (2-3 nm) ZnCdS shell and long-chain oleate ligands that were previously shown to prevent any measurable exciton diffusion.[1] The imaged PL spot is shown if the left panel of Fig. S6. The width of this distribution represents the convolved excitation and collection PSFs. With this characterization in hand, the imaged PL emission spot represents the convolution of the excitation and collection PSFs as well as additional broadening due to diffusion. This broadening can be used to extract a diffusivity, $D = 0.03 \pm 0.01$ cm$^2$/s as described in the main text.

## VII. PL Spectra under Extended Vacuum and Cooling

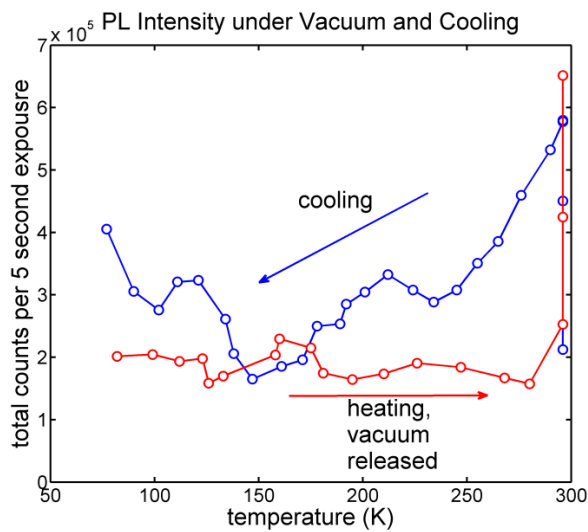

**Figure S7** *Integrated photoluminescence intensity upon cooling (blue) to and heating (red) from liquid nitrogen temperature. The cooling and heating cycle took approximately one hour.*

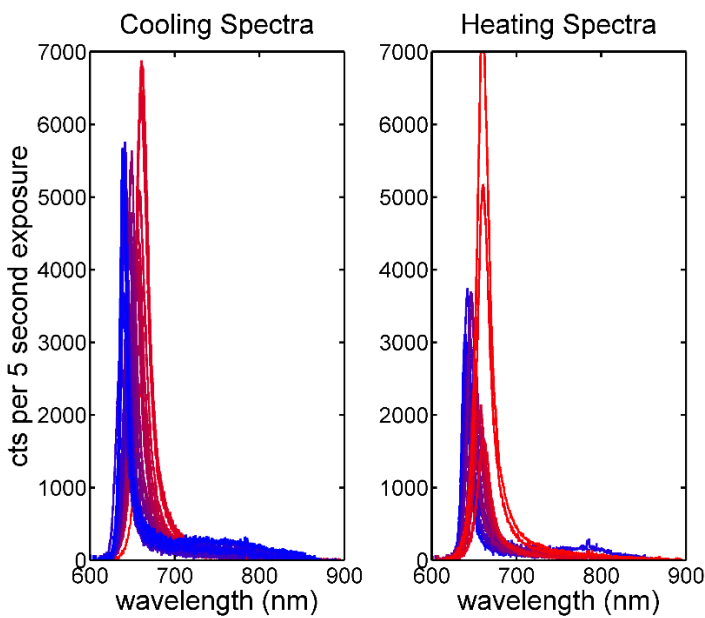

**Figure S8** *Photoluminescence spectra collected upon cooling (left) to and heating (right) from liquid nitrogen spectra. The color gradation indicates temperature (room temperature is red, 77 K is blue). The cooling and heating cycle took approximately one hour.*

To assess the effect of vacuum and low temperature on the polymer-capped samples, the samples were cooled to 77 K and heated to room temperature under vacuum. Upon cooling and concurrent exposure to vacuum, the total PL intensity decreased. Simultaneously, the exciton

S8

emission blue shifted and trap state emission became more prominent as less thermal energy was available to promote trapped excitons to the band edge.[2] Upon heating, the spectrum red shifted and the trap state emission became less prominent again. Notably, the PL intensity recovered only somewhat until the cryostat was returned to atmospheric pressure, at which point the PL intensity completely recovered.

## VIII. Fitting Low Temperature PL Spectra

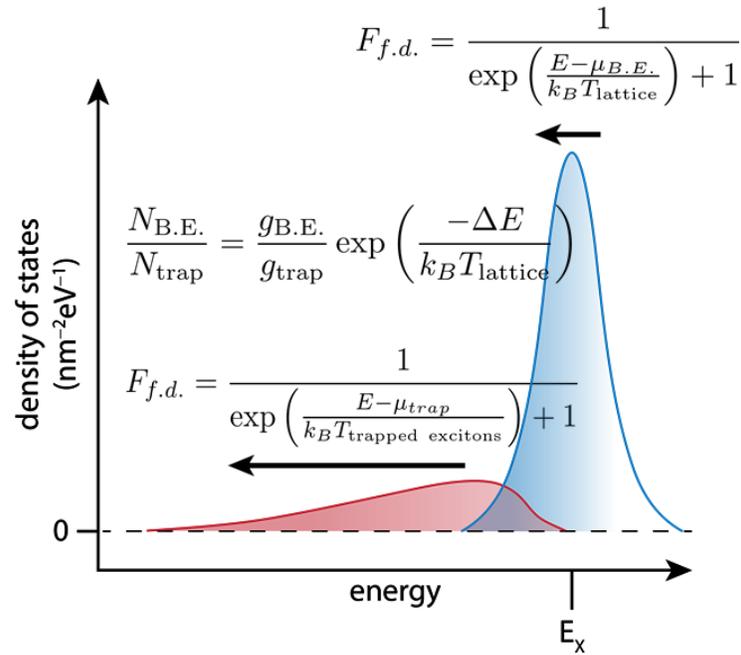

**Figure S9** *Schematic illustration of the two sub populations that contribute to low-temperature PL spectra.*

Low-temperature PL spectra exhibit contributions from band-edge excitons (blue) and trapped exciton states (red). Equilibration between these states is thermodynamically driven; their relative occupation is determined by the states' degeneracies and the energy difference between the states. Band-edge and trapped excitons can also cool within their respective distributions. The two densities of states are occupied by a Fermi-Dirac distribution. Cooling within the trap site density of states is diffusion-mediated and kinetically limited. The "effective temperature" of the trapped exciton population is much higher than the lattice temperature as the population does not have sufficient time to reach equilibrium. The PL spectra are well fit using the trapped exciton and band-edge exciton states' degeneracies and radiative rates.

# IX. Monte Carlo Modelling of Experimental Data

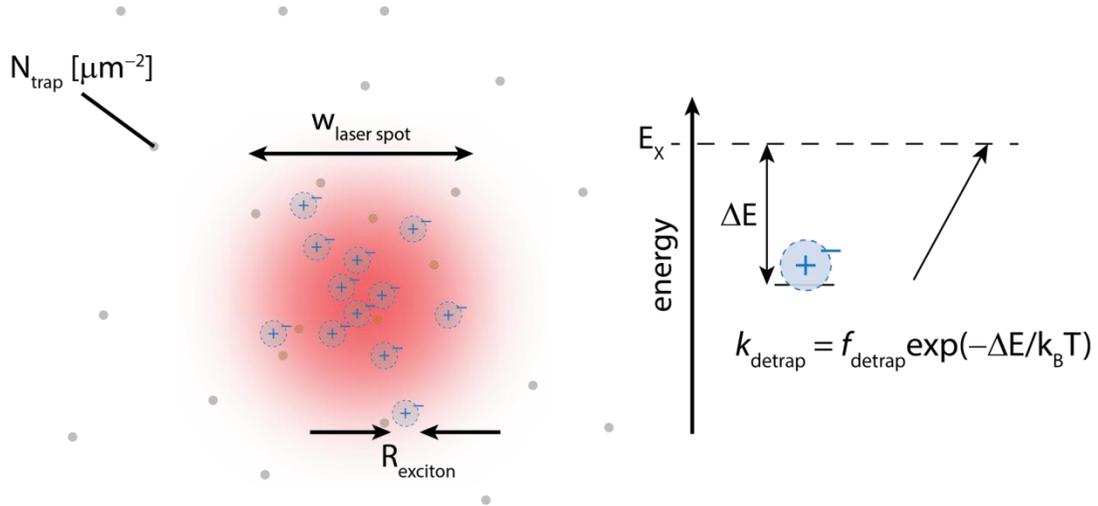

**Figure S10** *Left: Simulation schematic. Explicit traps are randomly initialized with density, $N_{trap}$. Excitons are randomly initialized under the Gaussian intensity profile of the excitation laser. Excitons interact when their centers are separated by $2*R_{exciton}$. Right: Schematic of the detrapping process. Traps have energies drawn from an exponential tail to the band edge state as detailed below. Trapped excitons detrap probabilistically with rate $f_{\text{detrap}} \exp\left(-\frac{\Delta E}{k_B T}\right)$.*

Steady-State Quantum Yield Implementation
The steady-state quantum yield for a given generation rate, $R_{gen}$, is modelled as follows:
1) The simulation box is populated with traps with a prescribed area density, $N_{trap}$, and an exponentially decaying energetic distribution, $\rho \propto \exp[-\alpha(E - E_{band\,edge})]$, where the parameter $\alpha$ is extracted from low temperature PL spectra.
2) The simulation box is populated with free excitons with rate $R_{gen}$ following the Gaussian intensity profile of the excitation laser.
3) Free excitons diffuse with the band edge diffusivity and decay with the band edge radiative rate.
   a. If a free exciton encounters an empty trap, it traps.
   b. If a free exciton encounters a trapped exciton, it annihilates.
   c. Trapped excitons detrap with rate $f_{\text{detrap}} \exp\left(-\frac{\Delta E}{k_B T}\right)$.
4) The number of emitted photons is divided by the number of generated excitons to extract the quantum yield.

Transient TCSPC Implementation
The transient simulations are performed for fixed initial exciton density, $N(t = 0)$. These simulations yield simulated diffusivities, $D$, and transient PL traces.

1) The simulation box is populated with traps with a prescribed area density, $N_{trap}$, and an exponentially decaying distribution with energy $\rho \propto \exp[-\alpha(E - E_{band\,edge})]$, where the parameter $\alpha$ is extracted from low temperature PL spectra.
2) The simulation box is populated with $N(0)$ excitons following the Gaussian intensity profile of the excitation laser.
3) Free excitons diffuse with the band edge diffusivity and decay with the band edge radiative rate.
   a. If a free exciton encounters an empty trap, it traps.
   b. If a free exciton encounters a trapped exciton, it annihilates.
   c. Trapped excitons detrap with rate $f_{detrap} \exp\left(-\frac{\Delta E}{k_B T}\right)$.
4) Whenever a free exciton decays radiatively, its emission time and spatial location are recorded. The emission times are histogrammed to model transient PL decay curves. The locations can be plotted to reproduce diffusion imaging experiments.

| | | Quartz | Sapphire | Strontium Titanate |
|---|---|---|---|---|
| $D_{band\,edge}$ | [cm$^2$s$^{-1}$] | | 5.0 | |
| Trap Density | [μm$^{-2}$] | | 1000 | |
| $f_{detrap}$ | [s$^{-1}$] | | 5×10$^{13}$ | |
| $\alpha$ | [eV$^{-1}$] | 6.0 | 7.5 | 11 |
| $R$ | [nm] | 4 | 0.6 | 0.4 |

**Table S1** *Parameters used in the Monte Carlo modelling of exciton diffusion, trapping, and annihilation. The band edge diffusivity ($D_{band\,edge}$), trap density, and detrapping attempt frequency ($f_{detrap}$) were the same when modelling each substrate. The trap state exponential tail parameter ($\alpha$) and annihilation radius ($R$) each depended on the supporting substrate in the system being modelled.*

Transient Monte Carlo Model Predictions
*Diffusivity Simulation*

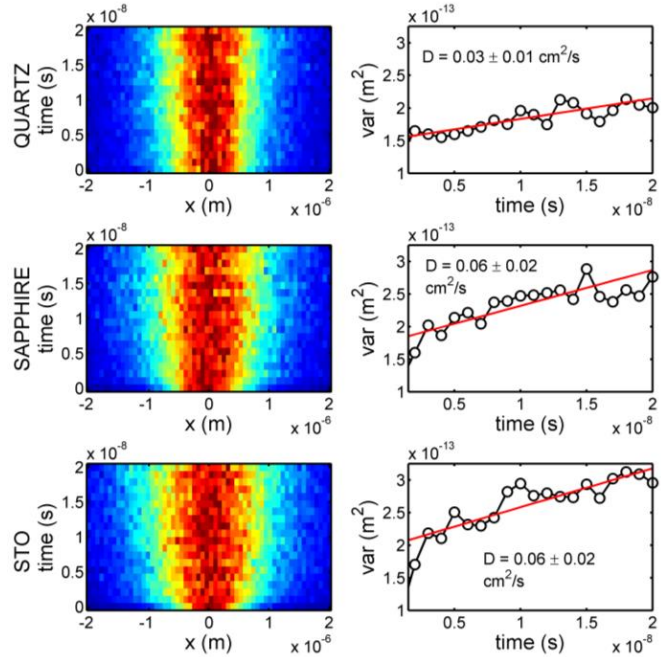

**Figure S11** *Simulated transient PL microscopy predictions. The Left column plots the simulated exciton diffusion imaging experiment. The simulated excitation density was one exciton per µm². In each 2D plot, time evolves along the vertical axis. Horizontal cuts represent normalized exciton spatial distributions diffusively broadening in time. The right column plots the variance of the Gaussian spatial profile as a function of time. The fitted diffusivities are also indicated. The model predictions are plotted for MoS$_2$ on quartz (top row), sapphire (middle row), and STO (bottom row).*

Our Monte Carlo model with explicit excitons and traps accurately reproduces the experimental diffusion imaging data. The band edge diffusivity used for each model is 5 cm$^2$/s. As seen in the main text, the measured diffusivity is orders of magnitude lower than the band edge diffusivity and matches what we measured experimentally. The measured diffusivity is smaller because excitons spend a large portion of their lifetime in immobile traps.

*Simulated Decay Dynamics and Transient Annihilation*

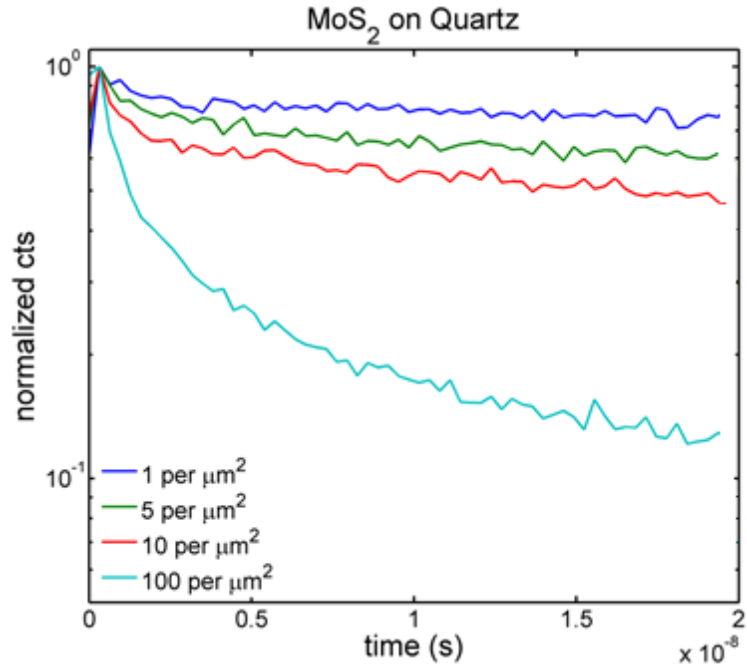

**Figure S12** *Simulated transient PL decay traces. The Monte Carlo simulation predictions for fluences of 1 (blue), 5 (green), 10 (red), and 100 (cyan) excitons per µm$^2$ are normalized and plotted. The simulated sample is MoS$_2$ on quartz.*

Our Monte Carlo model with explicit excitons and traps accurately reproduces the experimental PL decay dynamics. The band edge radiative rate is set to 900 ps. As seen in the main text, the predicted effective radiative lifetime is orders of magnitude longer (10s of ns) as seen in experiment. The measured effective radiative lifetime is longer due to the equilibration between band edge excitons and trapped excitons. The transient simulation also accurately captures exciton annihilation, which is most prominent on quartz. There is evidence of annihilation in simulated fluences as low as 5 excitons per $\mu$m$^2$.

**References:**
(1) Akselrod, G. M.; Prins, F.; Poulikakos, L. V.; Lee, E. M. Y.; Weidman, M. C.; Mork, A. J.; Willard, A. P.; Bulovic, V.; Tisdale, W. A. "Subdiffusive Exciton Transport in Quantum Dot Solids" *Nano Lett.* **2014,** *14*.
(2) Goodman, A. J.; Willard, A. P.; Tisdale, W. A. "Exciton Trapping Is Responsible for The Long Apparent Lifetime in Acid-Treated MoS$_2$" *Phys. Rev. B* **2017,** *96*, 121404(R).